\documentclass[useAMS,usenatbib]{mn2e}
\usepackage{epsfig}
\title[Colours of isolated galaxies selected from the Two-Micron All-Sky Survey]
  {Colours of isolated galaxies selected from the Two-Micron All-Sky Survey}

  \author[O. Melnyk et al.]
  { O.~Melnyk$^{1,2,3}$, S.~Mitronova$^{4}$, V. Karachentseva$^{3}$\\
$^1$ Institut d'Astrophysique et de G\'eophysique,
     Universit\'e de Li\`ege, 4000 Li\`ege, Belgium  \\
$^2$Astronomical Observatory, Kyiv National University, 3 Observatorna St., 04053 Kyiv, Ukraine\\
$^3$Main Astronomical Observatory, Academy of Sciences of Ukraine, 27 Akademika Zabolotnoho St., 03680 Kyiv, Ukraine\\
 $^4$ Special Astrophysical Observatory of the Russian Academy of Sciences, Nizhnij Arkhyz, KChR, 369167, Russia\\
}
\date{Released 2013 Xxxxx XX}

\pagerange{\pageref{firstpage}--\pageref{lastpage}} \pubyear{2013}

\def\LaTeX{L\kern-.36em\raise.3ex\hbox{a}\kern-.15em
    T\kern-.1667em\lower.7ex\hbox{E}\kern-.125emX}

\begin{document}

\label{firstpage}

\maketitle

\begin{abstract}
The properties of isolated galaxies are mainly driven by intrinsic evolution and not by the external influence of their environments. 
Therefore the isolated galaxy sample may be considered as a reference sample to study different environmental effects. We made detailed comparisons between 
the near-infrared (2MASS) and optical (SDSS) colours of the 2MASS XSC selected sample of isolated 2MIG galaxies and with other objects 
from the wide range of denser environments (field, groups/clusters, triplets and pairs). We found that early type galaxies show similar $(J-H)_{rest}$ and 
$(g-r)_{rest}$ colours practically in all types of environments. Exception is for the massive 
early type galaxies located in compact pairs with high velocity difference ($dV\sim$180 km/s), which are significantly redder and brighter than isolated galaxies. 
We assume that these pairs are located in the centres of more populated groups and clusters. In general, galaxies in 
groups and pairs of spiral and late morphological types have redder near-infrared colours $(J-H)_{rest}$ than 2MIG 
isolated galaxies. The $(g-r)_{rest}$ colours of galaxies in groups and pairs with high velocity difference are 
also significantly redder than corresponding colours of 2MIG galaxies. On the contrary, the members of most compact pairs ($dV\sim$50 km/s, $R\sim$30 kpc) show the 
same $(g-r)_{rest}$ colour and even tend to be bluer than 2MIG galaxies. In summary, our results show that the colours of galaxies are strongly depend on the 
external factors.

\end{abstract}


\section{Introduction}

The influence of internal factors and environment on galaxy evolution has been the subject of detailed studies in recent years (Blanton et al. 2005, 
van der Wel 2008, Hoyle et al. 2012, etc). From these studies it has become apparent that the process leading to the evolution of a galaxy from one 
morphological type to another are similar in all environments, but they are more intensive in dense regions (see review by Blanton \& Moustakas 2009 
and references therein). The results from correlation studies between the main galaxy parameters such as colour, stellar mass and morphology have shown 
that the environmental dependence on colour is stronger than for morphology (Bamford et al. 2009). Deng et al. (2012) concluded 
that galaxy colour depends more on environment than galaxy luminosity.

To study the influence of the environment on galaxy properties (such as colours, morphology, luminosity etc.), it is necessary to have reference samples. 
The authors of environmental studies (see references below) used the various kinds of reference samples which could be roughly grouped into three categories:
(i) the galaxies from all environments (see for example Blanton et al. 2005, van der Wel 2008); (ii) the galaxies in the void regions (Rojas et al. 2004, Hoyle et al. 
2005, 2012, Sorrentino et al. 2006, Patiri et al. 2006, von Benda-Beckmann \& Muller (2008), Kreckel et al. 2011, 2012) and (iii) the galaxies selected by use of an 
isolation criterion (Varela et al. 2004, Vavilova et al. 2009, Fernandez Lorenzo et al. 2012; Trinh et al. 2013). Note that in certain cases the control sample 
consists of galaxies which have similar properties to the investigated sample (e.g., Patton et al. 2011).

By definition, the isolated galaxies are objects that have not been appreciably affected by their closest neighbours over the past few Gyr so their 
properties are mainly driven by intrinsic evolution and not by the external influence of their environment (see Karachentseva 1980, Verley et al. 
2007, Karachentseva et al. 2010 and references therein). Unlike ``field'' or ``non-clustered'' galaxies, which are simply objects located outside 
of groups and clusters (de Vaucouleurs 1971, Turner \& Gott 1977, Makarov \& Karachentsev 2011, Karachentsev et al. 2012), the isolated galaxies are 
the objects selected using a special ``isolation'' criterion (examples of the different approaches can be found in the works by Karachentseva 1973, 
Varela et al. 2004, Allam et al. 2005,  Elyiv et al. 2009, Karachentseva et al. 2010, Karachentseva et al. 2010a, Hernandez-Toledo et al. 2010, 
Karachentsev et al. 2011).

The most studied sample of isolated galaxies is the Catalogue of Isolated Galaxies (Karachentseva 1973, KIG) \footnote{The catalogue is known as 
KIG in Russian transcription but the abbreviation CIG is also widely used.}, which includes 1050 objects with  $m \leq 15.7$ and $\delta>-3^{\circ}$, 
i.e. $\sim$4\% in the CGCG catalogue (Zwicky et al. 1961-1968). The largest contribution to the analysis of  the KIG properties was from the AMIGA team, 
with their most prominent results in Lisenfeld et al. (2007, 2011), Durbala et al. (2008),  Leon et al. (2008), Sabater  et al. (2008, 2012).  
Recently, Fernandez Lorenzo et al. (2012) considered the colour properties of KIGs and showed that the isolated spiral galaxies are redder than 
objects in close pairs. At the same time the authors did not find clear differences between the colours of isolated galaxies and galaxies in group/field environments. 
Moreover, isolated and non-isolated early type galaxies showed a similar $g-r$.

The main aim of the present work is to consider the colour properties of the 2MASS Isolated Galaxies (2MIG; Karachentseva et al. 2010)  in comparison with galaxies 
in other i.e. denser (field, pair, group/cluster) environments. In Section 2 we present the sample of 2MIG galaxies and the selection criteria. 
In Section 3 we compare the colours of 2MIG galaxies which have faint companions and without them. We also compare the 2MIG colours with those of control samples 
randomly selected from the field, groups and clusters. The results of a colour comparison between the 2MIG galaxies and those in  group/cluster and pair/triplet 
environments is given in Section 4. In Section 5 we present a short
discussion about obtained results, and we close with our main
conclusions in Section 6.

\section {The 2MIG isolated galaxies}

The 2MIG entire sky catalogue of isolated galaxies (Karachentseva et al. 2010) was selected from 1.6 million objects of the 
Two Micron All-Sky Survey Extended Source Catalog ( 2MASS XSC; Jarrett et al. 2000).  Selection of objects was performed twice: 
(i) automatically according to the original criteria of isolation by Karachentseva (1973) adapted to the 2MASS data:

\begin{eqnarray} 
X_{1i}/a_i\geq s = 30
\label{lpar:eq}
\end{eqnarray}

and 

\begin{eqnarray}
4\geq a_i/a_1 \geq 1/4,
\label{lpar:eq}
\end{eqnarray}

where subscripts “1” and “i” refer to the fixed galaxy and its neighbours, respectively. According to these criteria, 
a galaxy with a standard angular diameter $a_1$ is considered isolated if its angular separation $X_{1i}$ from all its neighbours 
with “significant” angular diameters $a_i$ inside interval (2) is equal to or exceeds 30$a_{i}$.

(ii) visual inspecting of the images (DSS1,2\footnote{Digital Sky Survey: http://archive.eso.org/dss/dss} and 
SDSS\footnote{Sloan Digital Sky Survey: http://www.sdss.org})  of all isolated candidates to identify visible neighbours. 
Visual inspection allowed us to eliminate physically multiple systems with blue neighbours which  were missed in the 
automatic selection (see details in Karachentseva et al. 2010).  Finally, the 2MIG catalogue consists of 3227 galaxies brighter 
than $K_s$ = 12 mag and with angular diameters $a_{Ks}\geq $ 30$''$. In total we found approximately 6\% of isolated galaxies 
among the extended sources of 2MASS XSC survey.

There have been several works concerning the 2MIG catalogue:  
Karachentseva et al. (2011) reported the results of the search for the faint companions around the isolated galaxies;
Kudrya et al. (2011)  considered the statistical relations between different observational characteristics, and 
Kudrya \& Karachentseva (2012)  constructed the Tully-Fisher relations for the 2MIG galaxies. Coziol et al. (2011) 
made an extensive analysis of AGN impact in the 2MIG sample, and 
Anderson et al. (2013) considered the X-ray properties of 2MIGs with ROSAT data.

We have updated the data about the radial velocities of the 2MIG galaxies using the NED\footnote{NASA/IPAC Extragalactic Database: 
\\ http://ned.ipac.caltech.edu} and HyperLeda\footnote{Database for physics of galaxies: http://leda.univ-lyon1.fr/} 
(Paturel et al. 2003) databases. This has led to an increase in the number of known velocities from 2328 to 2573. 
In the current paper we consider only those objects with $V_h>$500 km/s. 
We have also revised the morphological types of galaxies according to the SDSS images and to the types given by Coziol et al. (2011). 
The authors of the latter paper took into account not only the visually defined types of 2MIGs but also their star formation history 
based on spectra analysis. This led to a further 232 galaxies in the 2MIG sample having new morphological types. In this paper, 
like in Karachentseva et al. (2010), we have adopted the scale: -2 -- E, 0 -- S0, 1 -- S0/a, 2 -- Sab, 3 -- Sb, 4 -- Sbc, 5 -- Sc, 
6 -- Scd, 7 -- Sd, 8 -- Sdm, 9 -- Im, 10 -- Ir. 

\begin{figure} 
\includegraphics[width=8.5cm,trim=15 15 20 20,clip]{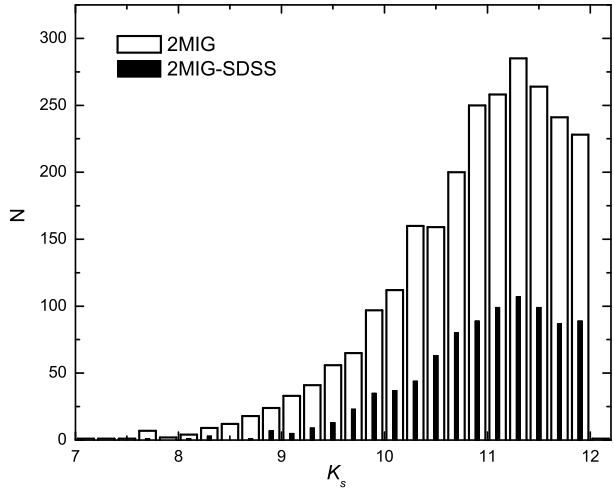}
\caption{The $K_s$ magnitude distribution for 2MIG (N=2541) and 2MIG-SDSS (N=912) galaxies.}
         \label{Fig1}
      \end{figure}

\begin{figure}
\includegraphics[width=8.5cm,trim=15 15 20 20,clip]{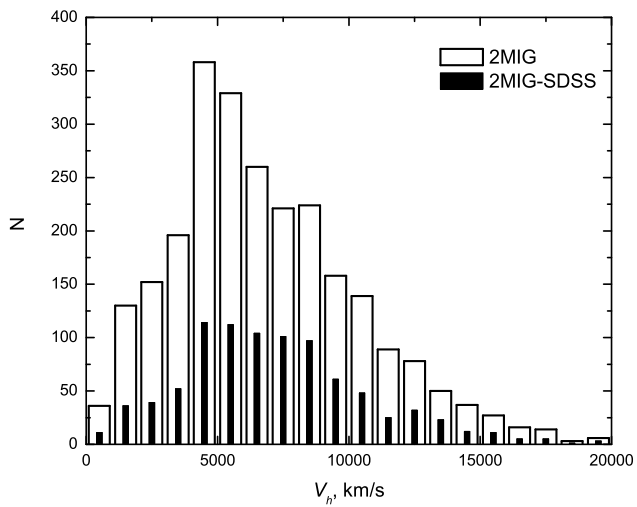}
\caption{The radial velocity distribution for 2MIG (N=2541) and 2MIG-SDSS (N=912) galaxies.}
\label{2}
\end{figure}

\begin{figure} 
\includegraphics[width=8.5cm,trim=15 15 20 20,clip]{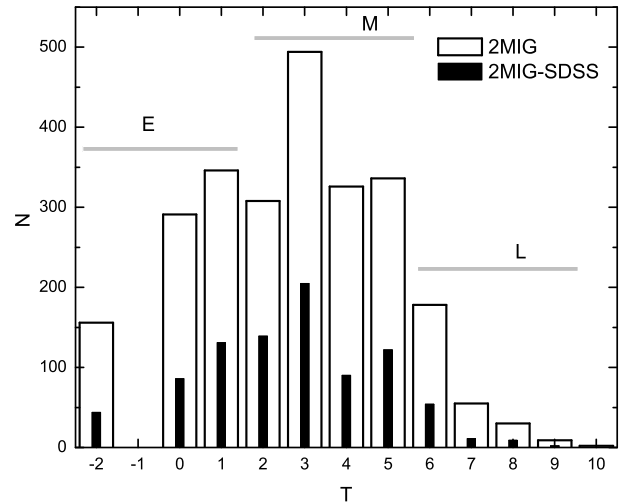}
\caption{The morphological type distribution for 2MIG (N=2541) and 2MIG-SDSS (N=912) galaxies. The united morphological types E, M and L are marked with grey lines.}
         \label{3}
      \end{figure}

In this work we have consider two galaxy samples: the 2MIG and 2MIG-SDSS galaxies having 2MASS and SDSS magnitudes, respectively.  
The 2MIG sample consists of galaxies which have all three near-infrared $J$, $H$, $K_s$ 2MASS XSC  magnitudes corresponding to 
isophote $K=$ 20 mag/arcsec$^{2}$. In  the 2MIG-SDSS sample we selected galaxies having $g$, $r$, $i$ magnitudes from the 
SDSS-III survey (DR8; Aihara et al. 2011). 

In Figs. 1, 2 and 3 we present the distributions of apparent magnitude $K_s$, radial velocity $V_h$ and morphological type T, 
for the 2MIG (N=2541) and 2MIG-SDSS (N=912) galaxies, respectively. It is seen from Fig. 1 that the 2MIG-SDSS sample follows good 
the 2MIG $K_s$ distribution (similar distributions we observe in $J$ and $H$ bands). The shapes of $V_h$ (Fig. 2) and T (Fig. 3) 
distributions for  2MIG-SDSS are also similar to the corresponding 2MIG results. The isolated galaxies present the whole 
scale of morphological types with notable maximum on the middle-spirals. The lack of the late type galaxies could potentially 
be explained by the $K_s$-band selection. In the work we considered the united morphological types: Early types E --
-2, 0, 1, Middle types M -- 2-5 and Late types L -- 6-10 (see grey lines in Fig. 3).  

For all galaxies considered in this paper we have calculated rest-frame colours as:

\begin{eqnarray} \nonumber
(M_{1}-M_{2})_{rest}=m_{1}-m_{2} - (K_{1}(z)-K_{2}(z)) - \\
 -(Q_{1}(z)-Q_{2}(z))-(A_{g1}-A_{g2})-A_{i12},
\label{lpar:eq}
\end{eqnarray}

\noindent where $m_{1}$ and $m_{2}$ are the apparent magnitudes transformed to the AB photometric system, 
$K_{1}(z)$ and $K_{2}(z)$, $Q_{1}(z)$ and $Q_{2}(z)$  are the $K$- and evolution corrections taken from Poggianti (1997), 
and $A_{g1}$ and $A_{g2}$ are the Galactic corrections for the first and second band, respectively. 
$A_{i12}$ is the correction for the internal absorption for morphological types T$>$1: according to Masters et al. (2003) for 2MASS 
colours and according to Masters et al. (2010) for SDSS colours. The Hubble constant is $H_0$=72 km/s/Mpc.

\section{Colours of 2MIG and galaxies in other environments}

\subsection{2MIGs with and without faint companions}

Karachentseva et al. (2011) showed that 125 2MIG galaxies have faint companions with close radial velocities 
of $dV <$ 500 km/s and projected separations of $R<$ 500 kpc relative to the 2MIG galaxies. These companions are not satisfied 
the condition (2), i.e. their diameters are too small in comparison with 2MIGs. However Karachentseva et al. (2011) showed that these companions have little
effect on the dynamic isolation of the 2MIG galaxies.

In Table 1 we compared the medians of ($J-H$)$_{rest}$ and ($g-r$)$_{rest}$ rest-frame colour indices according to formula (3) of 2MIG galaxies 
which have faint companions (2MIG$_{fc}$) and which do not have them (2MIG$_{nfc}$). We have considered the All samples and separately each united morphological type 
E, M and L (see explanation in Fig. 3). Table 2 represents the Kolmogorov-Smirnov (KS) probabilities 
that pairs of the samples are derived from the same parent population. We did not compare L type 
galaxies separately because of their paucity N=4 (3\%) in 2MIG$_{fc}$ sample. The part of L type galaxies in whole considered 
2MIG sample is near 11\% but it is not surprising that the earlier type galaxies tend to have more companions than the later types. The median values of the absolute magnitudes 
$M_{Ks}$ of 2MIG$_{fc}$ and 2MIG$_{nfc}$ are -22.18$^{+0.53}_{-0.49}$ and -21.98$^{+0.66}_{-0.56}$, respectively, which indicates that isolated galaxies
with faint companions are in general more massive than those without any neighbours, i.e. we see luminosity vs. morphology relation. The significance of the 
difference is confirmed by the KS probability given in Table 2. However if we consider E and M types separately, we do not see any significant difference 
in the magnitudes: -22.26$^{+0.55}_{-0.39}$ and -22.18$^{+0.53}_{-0.51}$ 
for E and M types of 2MIG$_{fc}$ sample, and -22.26$^{+0.62}_{-0.51}$ and -21.98$^{+0.60}_{-0.50}$ of 2MIG$_{nfc}$ sample, respectively.

Considering Tables 1 and 2 we conclude that E type galaxies with and without companions have completely same colours. Meanwhile M type galaxies 
have marginal difference in ($g-r$)$_{rest}$ colours assuming that isolated galaxies having the small companions seem slightly redder.  Independently from our
morphological classification, we have also compared the ($g-r$)$_{rest}$ colours of 2MIG$_{fc}$ and 2MIG$_{nfc}$ samples separating the early and late types 
by the value of the concentration index $C=R90/R50$\footnote{$R90$ and $R50$ are the radii containing 90\% and 50\% of the Petrosian galaxy light, respectively.}=2.60 
according to Strateva et al. (2001). Fig. 4 shows the correlation of $C$ vs. ($g-r$)$_{rest}$ with adopted morphological classification. From the last rows 
of Tables 1 and 2 we see that the result is completely agreed with the previous conclusion: the 2MIG$_{nfc}$ galaxies with $C<$2.6 are slightly 
bluer than 2MIG$_{fc}$ ones.

\begin{table}
\caption{The median values of colours in the quartile ranges for the united morphological types of 2MIGs with and without faint companions 
(2MIG$_{fc}$ and 2MIG$_{nfc}$, respectively), whole 2MIG and Control1/2 samples.} 
\tabcolsep 1 pt
\begin{tabular}{lccccccc} \hline
Samp-   &  \multicolumn{2}{c}{2MIG$_{fc}$} & \multicolumn{2}{c}{2MIG$_{nfc}$} &  & 2MIG & Control1/2 \\ 
le & N & $(J-H)_{rest}$  & N & $(J-H)_{rest}$ & N & \multicolumn{2}{c}{$(J-H)_{rest}$}   \\

\hline
All     & 125 & 0.21$^{+0.03}_{-0.04}$ & 2416 & 0.19$^{+0.03}_{-0.05}$ & 2541	& 0.20$^{+0.04}_{-0.05}$  &	 0.20$^{+0.04}_{-0.05}$   \\
E	&  42 & 0.23$^{+0.01}_{-0.02}$  & 754 & 0.22$^{+0.03}_{-0.02}$ & 796	& 0.22$^{+0.03}_{-0.02}$  &	 0.22$^{+0.03}_{-0.02}$    \\
	
M	&  79 & 0.20$^{+0.03}_{-0.06}$  & 1388 & 0.18$^{+0.04}_{-0.05}$  & 1467	&	0.18$^{+0.04}_{-0.05}$ &	   0.18$^{+0.04}_{-0.05}$    \\
	
L	&  4 & 0.04$^{+0.07}_{-0.01}$ & 284 & 0.13$^{+0.06}_{-0.07}$ & 328	&	0.13$^{+0.06}_{-0.07}$  &	  0.13$^{+0.05}_{-0.05}$   \\
\hline
 & N & ($g-r$)$_{rest}$  & N & ($g-r$)$_{rest}$ & N & \multicolumn{2}{c}{($g-r$)$_{rest}$}   \\
 \hline
All    & 81 & 0.75$^{+0.04}_{-0.08}$ & 831 & 0.71$^{+0.06}_{-0.08}$ & 912	&	0.72$^{+0.05}_{-0.08}$  &	 0.75$^{+0.06}_{-0.09}$   \\
E	& 25  & 0.77$^{+0.03}_{-0.02}$ & 240 & 0.78$^{+0.04}_{-0.04}$ & 265	&	0.78$^{+0.04}_{-0.04}$  &	 0.79$^{+0.05}_{-0.05}$    \\
	
M	& 53 & 0.73$^{+0.06}_{-0.08}$ & 508 & 0.69$^{+0.05}_{-0.07}$ & 561	&	0.69$^{+0.06}_{-0.07}$ &	   0.72$^{+0.06}_{-0.09}$    \\
	
L	& 3 & 0.61$^{+0.10}_{-0.18}$ & 83 & 0.54$^{+0.07}_{-0.07}$ & 86 &   	0.54$^{+0.07}_{-0.07}$  &	  0.69$^{+0.09}_{-0.15}$   \\
$C>$2.6 & 47 &  0.76$^{+0.04}_{-0.05}$  & 422 & 0.76$^{+0.05}_{-0.06}$ & 469* &   0.76$^{+0.05}_{-0.06}$ &       0.78$^{+0.05}_{-0.06}$ \\
$C<$2.6 & 34 &  0.72$^{+0.09}_{-0.07}$  & 409 & 0.66$^{+0.07}_{-0.08}$ & 443* &   0.66$^{+0.07}_{-0.09}$ &        0.68$^{+0.08}_{-0.10}$ \\
 \hline
\end{tabular}
 *The numbers are given for the 2MIG sample. For the Control2 sample N=557 and 355 for the $C>$2.6 and $C<$2.6, respectively. 
\end{table}

\begin{table}
\caption{The Kolmogorov-Smirnov probabilities that the pairs of samples 2MIG and Control1/2 are drawn from the same parent set.} 
\tabcolsep 3 pt
\begin{tabular}{lccccc} \hline
T  &  2MIG$_{fc}$/ & 2MIG$_{fc}$/ & 2MIG$_{fc}$/ & 2MIG/  & 2MIG/   \\
 &  2MIG$_{nfc}$ & 2MIG$_{nfc}$ & 2MIG$_{nfc}$ & Control1  & Control2   \\ 
 & $p_{M_{Ks}}$ & $p_{(J-H)_{rest}}$ & $p_{(g-r)_{rest}}$ & $p_{(J-H)_{rest}}$ & $p_{(g-r)_{rest}}$ \\
 \hline
 
All &  6.7$\times 10^{-3}$  &  4.1$\times 10^{-2}$ & 5.3$\times 10^{-2}$ & 9.4$\times 10^{-1}$ & 3.9$\times 10^{-7}$   \\
E   & 2.9$\times 10^{-1}$	& 7.7$\times 10^{-1}$ & 5.7$\times 10^{-1}$ & 9.9$\times 10^{-1}$  & 5.6$\times 10^{-1}$   \\
	
M   & 2.3$\times 10^{-1}$	& 4.3$\times 10^{-1}$ & 2.5$\times 10^{-2}$ & 3.3$\times 10^{-1}$	& 7.7$\times 10^{-6}$   \\
	
L   & - & - & - & 3.4$\times 10^{-1}$	& 3.2$\times 10^{-7}$  \\
$C>$2.6 & - & - & 9.9$\times 10^{-1}$ & - & 1.4$\times 10^{-3}$ \\
$C<$2.6 & - & - & 7.4$\times10^{-2}$ & - & 9.9$\times 10^{-2}$ \\
\hline
 \end{tabular}
\end{table}

\begin{figure} 
\includegraphics[width=8.5cm]{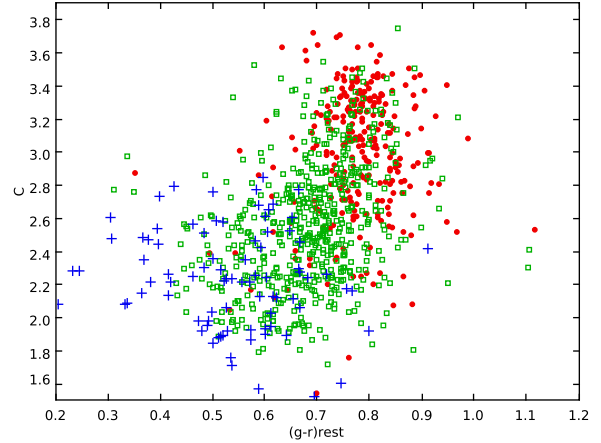}
\caption{Dependence of ($g-r$)$_{rest}$ colour on concentration index $C$ for 2MIG galaxies. Different symbols correspond to our visual united morphological types E, M 
and L which are decoded in Fig. 3.}
         \label{4}
     \end{figure}

\subsection{Galaxies in denser environments}

We have studied the colour properties of isolated galaxies in comparison with colours of galaxies located in denser regions (field, groups and clusters). 
To this end, we have compiled two samples of reference galaxies: the Control1 sample (which corresponds to the 2MIGs), 
and the Control2 sample (for the 2MIG-SDSS data). By analogy to the 2MIG criteria, our selection process was as follows: 
firstly, from the 2MASS XSC catalogue we selected candidate control objects which had $a_{K_s}\geq 30''$ and 4$\leq K_s\leq$12. 
Secondly, we downloaded the information about their morphological types and the radial velocities from the HyperLeda database. 
Considered control galaxy candidates have radial velocities V$<$25000 km/s like the galaxies from 2MIG sample.
The morphological HyperLeda's types for the control galaxy candidates were converted to the adopted scale. We independently 
inspected the images of 463 (18\% of the total number) randomly taken galaxies from Control1 sample and found that our and 
HyperLeda's morphological types of the Control1 galaxies coincide in 73\% of cases. However only 51 (11\%) galaxies have different united types 
E, M or L. We checked how this percent of mismatches influence on the values of colour indicies. In the 2MIG sample we changed the united morphological types 
of 280 galaxies (11\% from 2541) proportionally to found number of mismatches in the Control1 sample: 
39 L 2MIG types were changed to M, 76 M types -- to E, 120 E types -- to M,  34 M types -- to L and 11 E types to L. Finally, the median values of the colour indices 
of E, M and L united types of 2MIG galaxies are presented in Table 1. The corresponding medians and quartiles of ($J-H$)$_{rest}$ 
for the changed 2MIG sample are: 0.22$^{+0.03}_{-0.03}$ (E), 0.18$^{+0.04}_{-0.05}$ (M) and 0.14$^{+0.06}_{-0.07}$ (L). So we see that mismatches between
the HyperLeda and our morphological classification do not influence significantly on the final result.

Then, from the candidate control sample galaxies, we randomly selected the Control1 galaxies corresponding to 
each 2MIG galaxy according to the following criteria: (i)  $|M_{Ks,2MIG}-M_{Ks,Control1}|$=0.2 mag, (ii)  the morphological type of a Control1 
galaxy should belong to the same morphological subsample (E, M or L) as  the 2MIG galaxy and 
(iii) the same galaxies should not be present in the 2MIG and Control1 samples. From latter sample we also excluded the KIG and 
Hernandez-Toledo et al. (2010) isolated galaxies. Fig. 5 presents the distribution of $M_{Ks}$ for 2MIG and Control1 samples. 
According to the Kolmogorov-Smirnov (KS) test, the probability that the distributions of $M_{Ks}$ 
for the 2MIG and Control1 samples are identical, is greater than 0.99.

\begin{figure} 
\includegraphics[width=8.5cm,trim=0 20 0 0,clip]{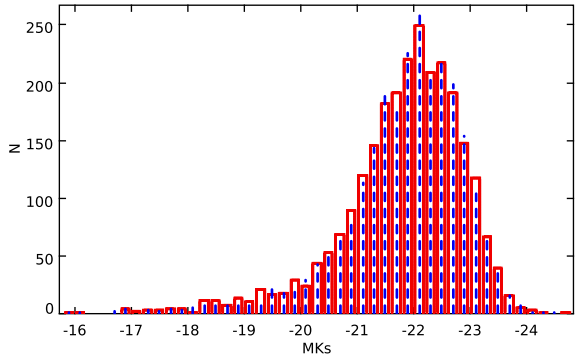}
\caption{Absolute magnitude $M_{Ks}$ distribution for 2MIG (open bar) and Control1 (dashed line bar) galaxies (N=2541).}
         \label{5}
     \end{figure}

\begin{figure*}
\tabcolsep 0 pt
\begin{tabular}{ccc}
\epsfig{file=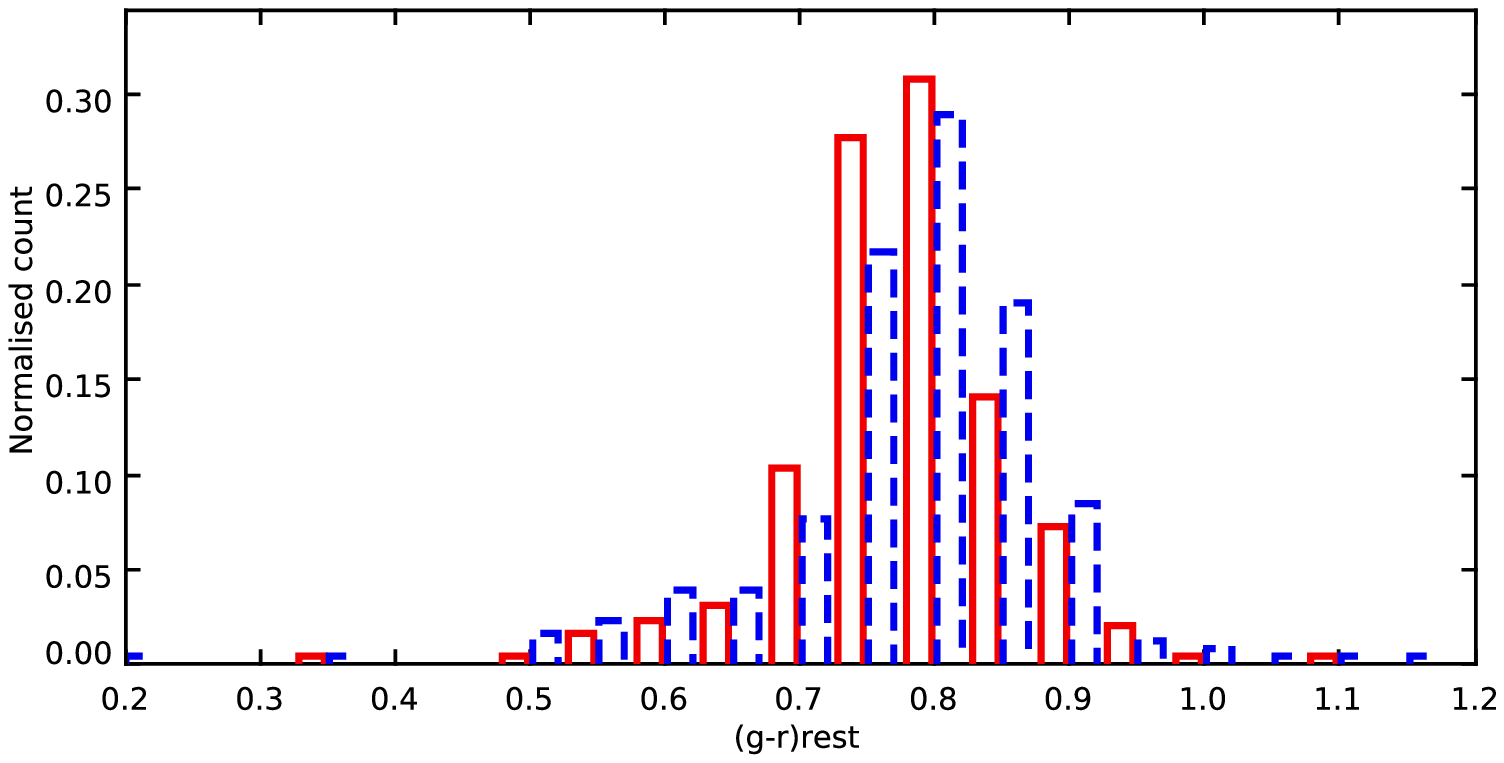,width=5.5cm,trim=10 10 25 0,clip} &
\epsfig{file=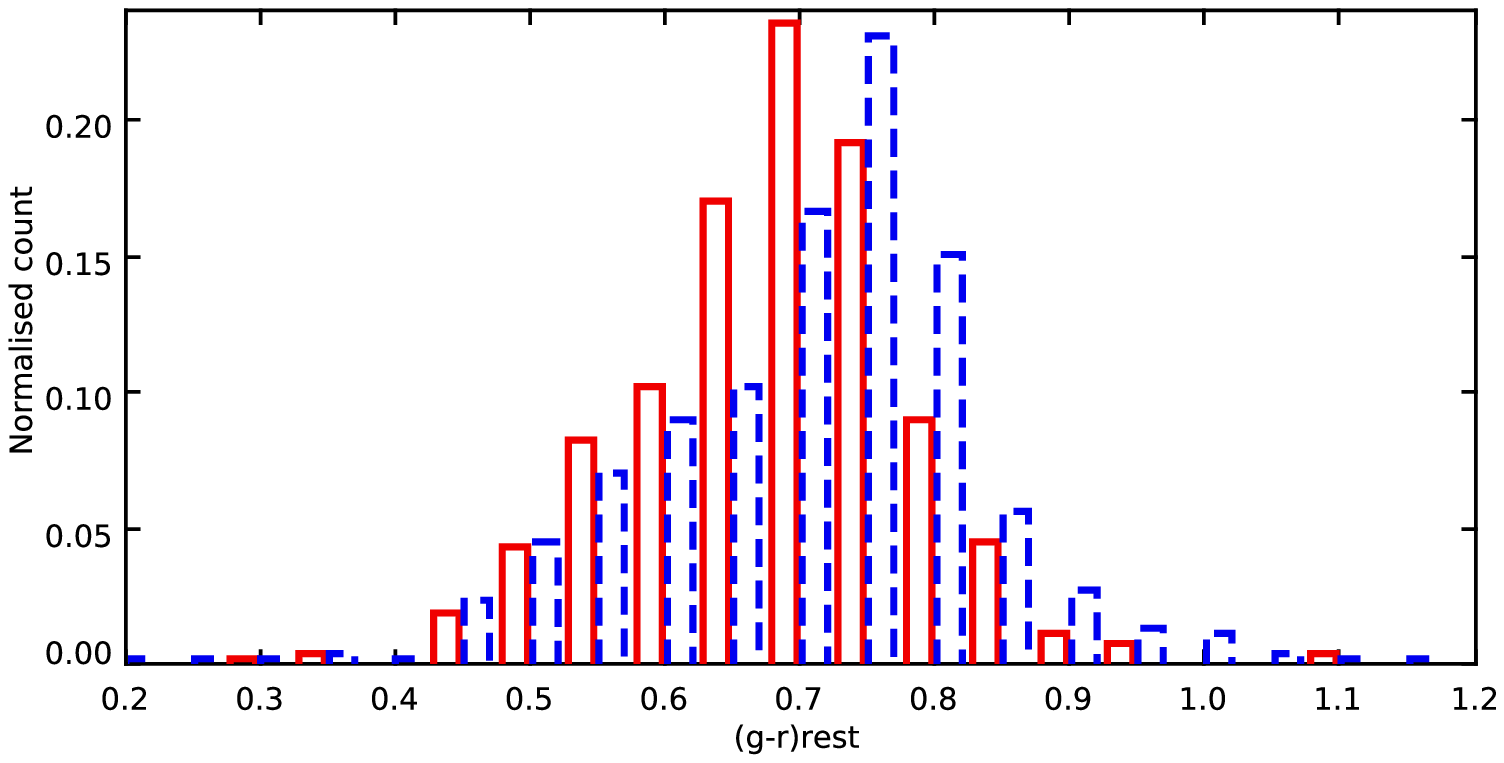,width=5.5cm,trim=10 10 25 0,clip} &
\epsfig{file=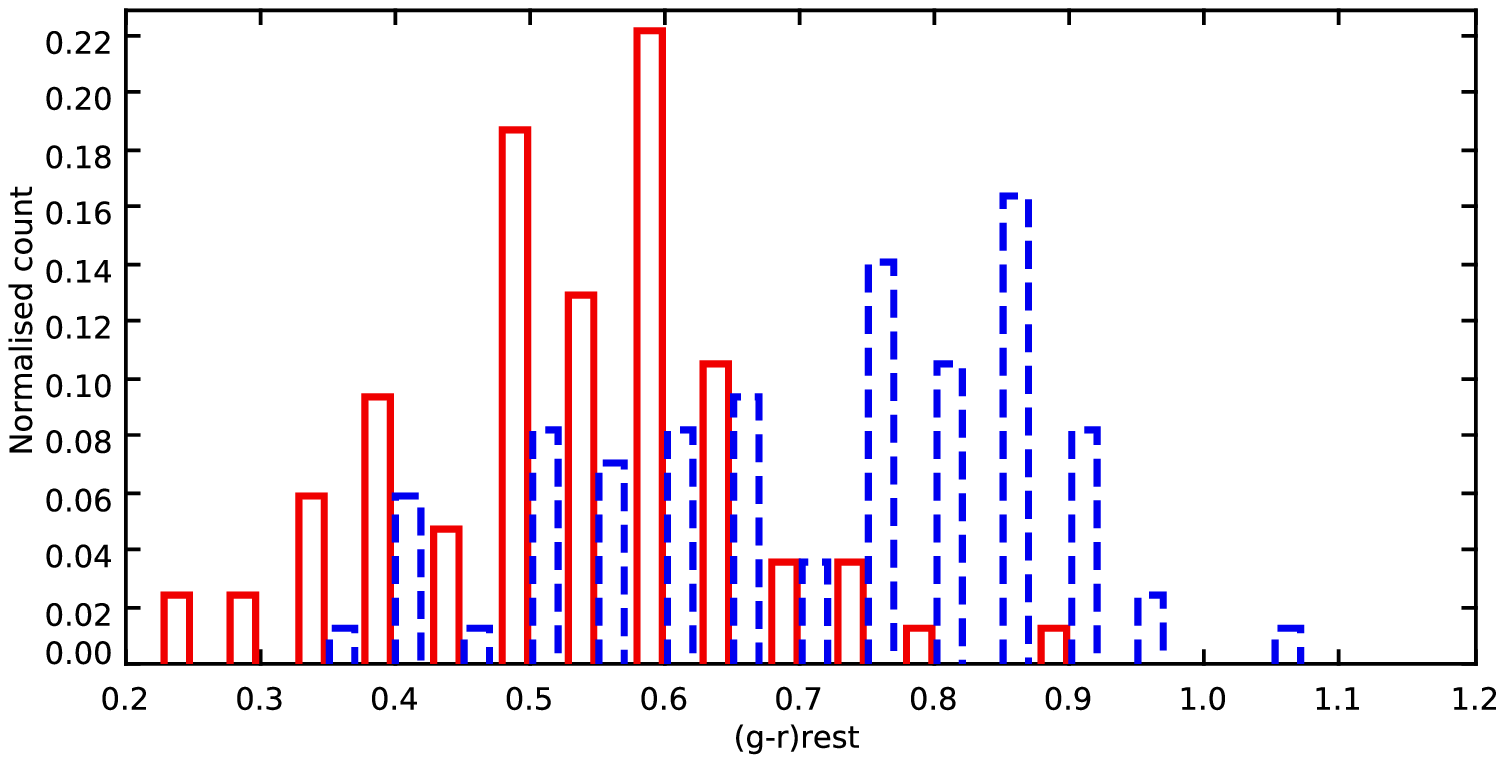,width=5.5cm,trim=10 10 25 0,clip} \\
E & M & L \\
\end{tabular}
\caption{Comparison of the colour distributions for E, M and L types of 2MIG-SDSS (open bar) and Control2 (dashed line bar) galaxies.} 
\label{6}
\end{figure*}

 We have also checked the environment of the Control1 galaxies to be sure that randomly taken galaxies are located in denser environment than 2MIGs. 
For our visually inspected representative subsample of 463 galaxies, we found that 22\% of them do not have any significant neighbour in 1 Mpc 
according to conditions (1) and (2), 24\% galaxies of the sample have 1 significant neighbour, and the rest 64\% of the sample has two and more neighbours.
So we can see that randomly selected galaxies in general inhabit denser environment.

Tables 1 represents the medians of rest-frame colour indices ($J-H$)$_{rest}$ of considering samples. Table 2 shows the KS probabilities that pairs of the samples 
are derived from the same parent population. We see that ($J-H$)$_{rest}$ colours 
of E, M and L types of the 2MIG galaxies do not differ significantly from those of the Control1 galaxies, which is also confirmed by the KS test. Probably we can explain 
this by 
$M_{Ks}$ selection of control galaxies -- condition (i).

We have also compared the SDSS colours of the 2MIG-SDSS and Control2 samples. To compile the Control2 sample we used a similar 
approach to that used for the Control1 sample. Firstly, we chose all candidates for control galaxies in the 2MASS XSC catalogue and 
HyperLeda using the same criteria as for the Control1 galaxies. Secondly, for all control galaxy candidates we checked 
the availability of the SDSS magnitudes. Finally, we chose randomly the Control2 galaxies among the candidates having SDSS magnitudes. Fig. 6 presents the 
differential distributions of $(g-r)_{rest}$ colours for E, M and L types of two considered samples.
From Tables 1 and 2 we see that E types of the 2MIG-SDSS and Control2 galaxies do not differ significantly, however $(g-r)_{rest}$ colours of 
M and L types of Control2 galaxies are significantly redder than 2MIGs. This discrepancy is not affected by dependence of
colour/morphological type from the radial velocity. In case of morphological separation by $C=$2.6, both samples ($C>$2.6 and $C<$2.6) of Control2 galaxies are 
significantly redder than corresponding 2MIG galaxies (see Tables 1 and 2). 

In general, our result agrees with the work by Fernandez Lorenzo et al. (2012) who also did not find any difference between $g-r$
colours of early type galaxies from KIG sample and the objects from a wide range of environments. 

The galaxies in voids are bluer and fainter than those in denser environments (see for example Rojas et al. 2004, Hoyle et al. 2005, 
Sorrentino et al. 2006). The same trend was found for void and cluster galaxies having the same absolute magnitudes (Hoyle et al. 2012). 
On the whole the void's population  are fainter and bluer than isolated galaxies but it could be a subsample of isolated galaxies (Elyiv et al. 2013). Therefore we 
conclude that our results of $(g-r)_{rest}$ colours support above findings. On the contrary,
Patiri et al. (2006) considered the red and blue population of void galaxies separately and did not find any remarkable 
differences between the colour distributions of the void and control sample of field galaxies.

\section{Colours of 2MIG in comparison with galaxies in much denser environments}

In this section we have made an attempt at comparing the colours of the 2MIG sample with those of galaxies in much denser environments. In the following subsections 
we considered three reference samples for this: (i) galaxies from the high density contrast groups selected from 2MASS XSC (HDCG; Crook et al. 2007), (ii) galaxies 
from pairs and triplets (KP+KT; Karachentsev 1987, Karachentseva et al. 1979, Karachentseva \& Karachentsev 2000), and (iii) galaxies from the compact 
pairs selected from the 2MASS XSC. By analogy to the previous section, at first we calculated the median values of the $M_{Ks}$ absolute magnitudes and rest-frame 
($J-H$)$_{rest}$ and $(g-r)_{rest}$ colour indices. Then we compared the corresponding values of (i), (ii) and (iii) samples with 2MIG sample using KS test 
(see Tables 3 and 4).

\subsection{Galaxies in groups}

Galaxies in the original HDCG catalogue have $K_s<$11.25 and $V_h<$10500 km/s. For the proper comparison with 2MIG sample, we defined 2MIG* sample following 
above criteria. Like in the previous section, we did not consider galaxies with $V_h<$500 km/s. We compared the colour indices of all galaxies from HDCG and 2MIG* samples
and separately the subamples of galaxies having E, M and L types. From Table 3 we see that the median values of $M_{Ks}$ for 2MIG* samples
are quite similar to HDCG samples, however the  ($J-H$)$_{rest}$ and $(g-r)_{rest}$ colours of the group members are a little redder for M and L types. The KS test 
confirms that the difference is significant (except for L types for ($J-H$)$_{rest}$, see Table 4). However if we separate morphology by the concentration index $C=$2.6, 
we see that the both HDCG samples $C>$2.6 and $C<$2.6 samples
are significantly redder than 2MIG* ones. The small KS probabilities for All samples can be naturally explained by the different contamination of E, M and L types in 
2MIG* (2MIG) and HDCG samples: we have 38\% (31\%) and 62\% of E type galaxies in the corresponding samples.

\subsection{Galaxies in pairs and triplets}

The KP+KT original systems were selected visually from POSS I and II\footnote{Second Palomar Observatory Sky Survey} images  by applying the isolation 
criteria without taking into account radial velocities. Therefore we updated their radial velocity data using the HyperLeda database and rejected the pairs 
and triplets having the velocity dispersion $S_v >$ 300 km/s. However if the triplet system has $S_v >$ 300 km/s, but for any two of its members $S_v$ 
was less than 300 km/s, we rejected the only one optical triplet member. We have to note that KP+KT systems contain relatively bright galaxies with $B<$15.7 mag 
in the same volume that 2MIG catalogue. Additionally, for the comparison with 2MIG sample, we take into account only galaxies 
which have $Ks<$12.  If we would consider the whole sample without any $Ks$ cut, then the median values of $(g-r)_{rest}$ colours for E, M and L types 
were 0.77$^{+0.04}_{-0.05}$, 0.62$^{+0.12}_{-0.16}$ and 0.43$^{+0.14}_{-0.17}$. So the M and L type galaxies became significantly bluer due to $B$-band 
sample selection in comparison with values given in Table 3. 

As can be seen from Tables 3, ($J-H$)$_{rest}$ colours of KP+KT galaxies of E, M and L types are slightly redder while the medians of $(g-r)_{rest}$ 
colours for all KP+KT subsamples are slightly bluer than those of 2MIGs. In general, the differences are not significant according to KS test. The exception is M-type
KP+KT galaxies, which are notably bluer than 2MIG M-types (Table 4). The total percent of E type galaxies in KP+KT sample is 47\% which is also higher 
than in 2MIG sample -- 31\%.

\begin{table*}
\caption{The median values of colours in the quartile ranges for the united morphological types of 2MIG, HDCG, KP+KT and CP1/2/3 galaxies.} 
\tabcolsep 2 pt
\begin{tabular}{lcccccccccccccc} \hline
T  & \multicolumn{2}{c}{2MIG} &  \multicolumn{2}{c}{2MIG*} & \multicolumn{2}{c}{HDCG} & \multicolumn{2}{c}{KP+KT} & \multicolumn{2}{c}{CP1} & \multicolumn{2}{c}{CP2} & \multicolumn{2}{c}{CP3} \\
 \hline
 & N & $M_{Ks}$ & N & $M_{Ks}$  & N & $M_{Ks}$  & N & $M_{Ks}$ & N & $M_{Ks}$ & N & $M_{Ks}$ & N & $M_{Ks}$ \\
  \hline
 All & 2541 & -21.99$^{+0.65}_{-0.56}$ & 1486 & -21.97$^{+0.59}_{-0.54}$ &6165 & -22.01$^{+0.69}_{-0.56}$ & 887 & -22.00$^{+0.75}_{-0.64}$ & 2351 & -22.29$^{+0.64}_{-0.59}$ & 1349 & -22.07$^{+0.65}_{-0.59}$ & 491 & -21.74$^{+0.70}_{-0.64}$\\
E & 796 & -22.26$^{+0.61}_{-0.51}$ & 561	& -22.14$^{+0.54}_{-0.50}$  & 3809 & -22.13$^{+0.64}_{-0.58}$ & 390 & -22.15$^{+0.67}_{-0.56}$  & 1363 & -22.47$^{+0.70}_{-0.56}$ & 736 & -22.27$^{+0.62}_{-0.58}$ & 246 & -21.96$^{+0.64}_{-0.68}$ \\
	
M & 1467 & -21.99$^{+0.59}_{-0.50}$ & 812	& -21.96$^{+0.56}_{-0.51}$ & 2037 &  -21.89$^{+0.63}_{-0.52}$ & 415 & -21.99$^{+0.77}_{-0.66}$ & 892 & -22.12$^{+0.64}_{-0.47}$ & 546 & -21.89$^{+0.63}_{-0.45}$ & 213 & -21.63$^{+0.66}_{-0.54}$ \\
	
L & 328	& -20.81$^{+1.21}_{-0.73}$ & 113	& -20.57$^{+1.30}_{-0.77}$  & 319 & -20.75$^{+1.08}_{-0.86}$ & 82 & -21.01$^{+1.32}_{-0.89}$ & 96 & -22.01$^{+1.09}_{-0.75}$ & 67 & -21.48$^{+0.86}_{-0.89}$ & 32 & -20.99$^{+0.91}_{-1.17}$ \\
 \hline
 & N & $(J-H)_{rest}$  & N & $(J-H)_{rest}$ & N & $(J-H)_{rest}$  & N & $(J-H)_{rest}$ & N & $(J-H)_{rest}$ & N & $(J-H)_{rest}$ & N & $(J-H)_{rest}$ \\
 \hline
 
All & 2541 & 0.20$^{+0.04}_{-0.05}$  & 1486   & 0.20$^{+0.03}_{-0.05}$ & 6165 & 0.20$^{+0.04}_{-0.05}$ & 887 & 0.21$^{+0.03}_{-0.04}$ & 2351 & 0.23$^{+0.04}_{-0.04}$ & 1349 & 0.23$^{+0.03}_{-0.05}$ & 491 & 0.22$^{+0.03}_{-0.06}$ \\
E & 796	& 0.22$^{+0.03}_{-0.02}$  & 561	& 0.22$^{+0.03}_{-0.02}$  & 3809 & 0.22$^{+0.03}_{-0.04}$ & 390 & 0.22$^{+0.03}_{-0.02}$ & 1363 & 0.26$^{+0.03}_{-0.03}$ & 736 & 0.25$^{+0.03}_{-0.03}$ & 246 & 0.25$^{+0.03}_{-0.03}$  \\
	
M & 1467 & 0.18$^{+0.04}_{-0.05}$ & 812	& 0.17$^{+0.04}_{-0.05}$ & 2037 &  0.18$^{+0.05}_{-0.06}$ & 415 & 0.19$^{+0.04}_{-0.04}$ & 892 & 0.19$^{+0.04}_{-0.05}$ & 546 & 0.18$^{+0.04}_{-0.05}$ & 213 & 0.17$^{+0.04}_{-0.05}$ \\
	
L & 328	&	0.13$^{+0.06}_{-0.07}$  & 113	& 0.12$^{+0.05}_{-0.06}$ & 319 & 0.14$^{+0.06}_{-0.08}$ & 82 & 0.15$^{+0.06}_{-0.08}$ & 96 & 0.19$^{+0.05}_{-0.06}$ & 67 & 0.16$^{+0.06}_{-0.04}$ & 32 & 0.16$^{+0.06}_{-0.04}$ \\
\hline
 & N & $(g-r)_{rest}$  & N & $(g-r)_{rest}$  & N & $(g-r)_{rest}$  & N & $(g-r)_{rest}$ & N & $(g-r)_{rest}$ & N & $(g-r)_{rest}$ & N & $(g-r)_{rest}$ \\
 \hline
 All & 912	&	0.72$^{+0.05}_{-0.08}$  & 512 & 0.73$^{+0.06}_{-0.07}$ & 2417 & 0.77$^{+0.04}_{-0.06}$ & 607 & 0.74$^{+0.06}_{-0.13}$ & 965 & 0.78$^{+0.04}_{-0.06}$ & 539 & 0.77$^{+0.05}_{-0.07}$ & 192 & 0.75$^{+0.06}_{-0.10}$ \\
E & 265	&	0.78$^{+0.04}_{-0.04}$  & 181	& 0.78$^{+0.05}_{-0.04}$  & 1549 & 0.78$^{+0.04}_{-0.04}$ & 284 & 0.77$^{+0.04}_{-0.04}$ & 544 & 0.80$^{+0.03}_{-0.04}$ & 278 & 0.79$^{+0.03}_{-0.04}$ & 92 & 0.78$^{+0.04}_{-0.05}$ \\
	
M & 561	&	0.69$^{+0.06}_{-0.07}$ & 293	& 0.70$^{+0.05}_{-0.06}$ & 755 &  0.73$^{+0.06}_{-0.08}$ & 269 & 0.68$^{+0.09}_{-0.12}$ & 380 & 0.73$^{+0.06}_{-0.07}$ & 232 & 0.71$^{+0.07}_{-0.09}$ & 88 & 0.71$^{+0.08}_{-0.13}$ \\
	
L & 86 &   	0.54$^{+0.07}_{-0.07}$  & 38	& 0.53$^{+0.08}_{-0.07}$  & 133 & 0.61$^{+0.08}_{-0.13}$ & 54 & 0.52$^{+0.11}_{-0.08}$ & 41 & 0.78$^{+0.10}_{-0.20}$ & 29 & 0.67$^{+0.19}_{-0.16}$ & 12 & 0.55$^{+0.21}_{-0.11}$\\
$C>$2.6 & 469 &   0.76$^{+0.05}_{-0.06}$ & 286 & 0.76$^{+0.04}_{-0.06}$ & 1793 & 0.78$^{+0.04}_{-0.04}$ & 387 & 0.77$^{+0.04}_{-0.06}$ & 720 & 0.79$^{+0.04}_{-0.04}$ & 372 & 0.78$^{+0.04}_{-0.05}$ & 720 & 0.78$^{+0.05}_{-0.06}$ \\
$C<$2.6 & 443 &   0.66$^{+0.07}_{-0.09}$ & 226 & 0.67$^{+0.07}_{-0.07}$ & 624 & 0.69$^{+0.09}_{-0.09}$ & 220 & 0.65$^{+0.07}_{-0.12}$ & 245 & 0.70$^{+0.08}_{-0.11}$ & 167 & 0.69$^{+0.09}_{-0.12}$ & 720 & 0.69$^{+0.09}_{-0.17}$ \\
 \hline
\end{tabular}
2MIG* sample is limited by $Ks<$11.25 and $Vh<$10500 km/s for the comparison with HDCG sample which was selected by these criteria.
\end{table*}

\begin{table*}
\caption{The Kolmogorov-Smirnov probabilities that the pairs of samples are drawn from the same parent set.} 
\tabcolsep 3 pt
\begin{tabular}{lccccccc} \hline
T  &  2MIG*/HDCG  & 2MIG/KP+KT  & 2MIG/CP1 & 2MIG/CP2 & 2MIG/CP3 & CP1/CP3 & KT+KT/CP3 \\ 
 & \multicolumn{7}{c}{$p_{M_{Ks}}$} \\
 \hline
 All & 7.5$\times 10^{-2}$ & 1.6$\times 10^{-1}$  & 4.9$\times 10^{-22}$ & 5.4$\times 10^{-2}$ & 2.8$\times 10^{-4}$  & 1.4$\times 10^{-20}$ & 1.1$\times 10^{-3}$\\
E & 1.0$\times 10^{-1}$ & 2.5$\times 10^{-1}$  &	4.2$\times 10^{-6}$ & 1.7$\times 10^{-1}$  & 9.5$\times 10^{-5}$  & 1.4$\times 10^{-9}$  & 6.8$\times 10^{-2}$\\
	
M & 4.2$\times 10^{-2}$& 3.6$\times10^{-2}$ & 2.7$\times10^{-3}$ & 1.9$\times10^{-2}$ & 1.2$\times10^{-5}$	& 9.9$\times 10^{-9}$ &	 8.4$\times 10^{-5}$   \\
	
L & 4.1$\times10^{-1}$ & 6.7$\times 10^{-2}$ & 1.6$\times 10^{-8}$  & 2.5$\times 10^{-3}$  & 3.7$\times 10^{-1}$ &	 1.8$\times 10^{-1}$ &	 7.0$\times 10^{-1}$\\
\hline
 & \multicolumn{7}{c}{$p_{(J-H)_{rest}}$} \\
 \hline
 All & 1.1$\times 10^{-6}$ & 3.2$\times 10^{-5}$  & 3.4$\times 10^{-23}$ & 3.9$\times 10^{-15}$ & 7.9$\times 10^{-11}$ & 1.9$\times 10^{-5}$ & 9.7$\times 10^{-5}$ \\
E & 4.5$\times 10^{-1}$ & 9.4$\times 10^{-1}$ & 3.1$\times 10^{-18}$ & 3.5$\times 10^{-11}$  &	2.6$\times 10^{-9}$ & 2.8$\times 10^{-3}$ & 1.9$\times 10^{-2}$  \\
	
M & 4.5$\times 10^{-2}$ & 2.2$\times10^{-2}$ & 2.7$\times 10^{-3}$ &	 2.0$\times 10^{-1}$  &	 2.8$\times 10^{-1}$ &	 2.6$\times 10^{-3}$ &	 5.2$\times 10^{-2}$ \\
	
L & 2.4$\times 10^{-1}$ & 3.5$\times 10^{-1}$	&	3.5$\times 10^{-5}$  &	 1.3$\times 10^{-3}$  &	 1.3$\times 10^{-3}$  &	 2.0$\times 10^{-1}$ &	 9.0$\times 10^{-1}$\\ 
 \hline
 & \multicolumn{7}{c}{$p_{(g-r)_{rest}}$} \\
 \hline
  All & 2.8$\times 10^{-13}$  & 2.7$\times 10^{-3}$ &  6.8$\times 10^{-20}$ & 2.4$\times 10^{-14}$ & 5.6$\times 10^{-4}$ & 1.7$\times 10^{-3}$ & 4.2$\times 10^{-1}$ \\
E & 8.7$\times 10^{-1}$ & 7.9$\times 10^{-1}$  &	5.5$\times 10^{-3}$ & 3.0$\times 10^{-1}$  & 8.9$\times 10^{-1}$ & 9.1$\times 10^{-2}$ & 4.1$\times 10^{-1}$ \\
	
M & 1.1$\times10^{-4}$	& 2.2$\times10^{-4}$	&	3.4$\times 10^{-8}$ &	 7.1$\times 10^{-3}$ &	 1.5$\times 10^{-1}$ &	 8.9$\times 10^{-2}$ &	 5.9$\times 10^{-1}$ \\
	
L & 1.8$\times10^{-2}$ & 8.4$\times 10^{-1}$ & 3.5$\times 10^{-7}$  &	 7.7$\times 10^{-4}$ &	  3.0$\times 10^{-1}$ &	 1.6$\times 10^{-1}$ &	 8.1$\times 10^{-1}$ \\
$C>$2.6	& 6.0$\times 10^{-3}$ & 1.0$\times 10^{-1}$	& 9.2$\times 10^{-14}$  & 4.4$\times 10^{-6}$ & 8.4$\times 10^{-2}$ & 2.2$\times 10^{-2}$ & 4.6$\times 10^{-1}$ \\
$C<$2.6	& 4.9$\times 10^{-3}$ & 3.4$\times 10^{-2}$ & 1.2$\times 10^{-4}$ & 3.9$\times 10^{-3}$ & 2.5$\times 10^{-2}$ & 1.6$\times 10^{-1}$  & 3.3$\times 10^{-1}$  \\
 \hline
\end{tabular}
\end{table*}

\subsection{Galaxies in the compact pairs}

To compare the colours of 2MIG galaxies with the galaxies selected from the densest regions, we have selected the compact galaxy pairs from 2MASS XSC catalogue. 
At first we extracted the galaxies with $Ks<$12 and $a_K<30''$ according to 2MIG selection criteria. Then, we found the closest neigbour for every galaxy 
from the sample taking into account only angular separation between galaxies. According to our procedure, the certain galaxy can be the member of the only one pair. 
By next step, we ranged the pairs from a minimal angular separation to maximal one and excluded the pairs with angular separation $<20''$. We found, that the great 
majority of pairs with such a small separation are double entitiles of the same object. By next step, we took 2200 most tight pairs and correlated 
the positions of their galaxies with objects in NED and LEDA databases. Finally, we rejected from the future consideration: 1) all pairs without radial velocities for at 
least one pair member, 2) the pairs with velocity difference $dV>$1000 km/s. Our final sample contains 1179 pairs with 2351 galaxies having $J$, $H$, $Ks$ 
magnitudes and 965 galaxies having SDSS magnitudes. We define 3 samples of compact pairs: CP1 with 
$dV<$1000 km/s and linear projected distance $R<$240 kpc (the whole sample), CP2 with $dV<$300 km/s and $R<$100 kpc, and CP3 
having $dV<$150 km/s and $R<$50 kpc (the most tight pairs).

From Tables 3 and 4 we see, that the members of pairs with the greater velocity difference (and also with greater projected radius) 
are brigher (i.e. more massive) and have redder colour than members of pairs with smaller velocity difference and also in comparison with isolated galaxies. 
In general, the $(J-H)_{rest}$ colours of CP1/2/3 members are redder than 2MIGs. However, the smaller distance between member of pair, the bluer the colour of pair member 
and this colour is more similar to 2MIG colour: i.e. members of CP1 pairs are much redder than 2MIGs while the colours of members of CP3 pairs is more 
similar to 2MIG colours. Especially this tendency is good visible for $(g-r)_{rest}$ colours. Additionally, from the last two columns of Tables 3 and 4 we can conclude 
that the colours of CP1 galaxies are significantly redder than CP3 galaxies while the colours of CP3 pair members are similar to KP+KT members. 

In the last raws of Tables 3 and 4
we also present the values of colours indices and KS probabilities for the morphological separation on the early and late population according 
to concentration index ($C=$2.6). As can be seen, both $C>$2.6 and $C<$2.6 galaxies from HDCG, CP1, CP2 samples are notably redder than 2MIG galaxies, 
however the differences between colours of pairs 2MIG/KP+KT and 2MIG/CP3 are less significant.

\section{Discussion of the main results}

\begin{figure} 
\includegraphics[width=9cm,trim=0 470 30 0,clip]{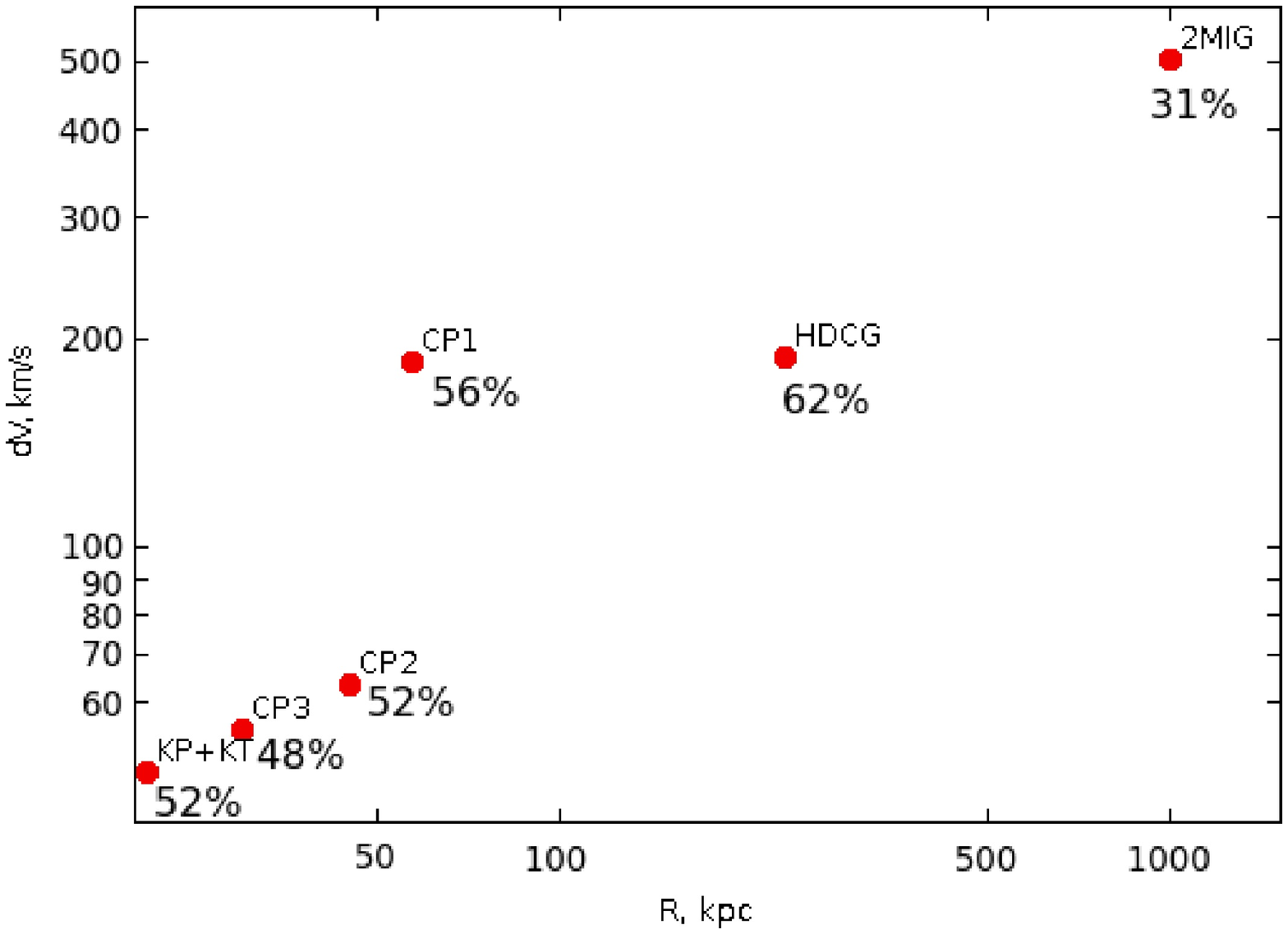}
\caption{Difference of the radial velocities for neighbouring galaxies vs. projected linear separation: median values. 
The percent of the early type galaxies in each sample are noted.}
        \label{7}
    \end{figure}

Fig. 7 shows the dependence of median values of the velocity difference $dV$ on projected linear radius $R$ between near neighbours for the considered samples. 
We show 
only the lowest limit of $dV$ and $R$ values for the 2MIG sample. We see that KP+KT, CP3 and CP2 pairs are the closest systems while the CP1 pairs and HDCG 
groups are characterized by the higher velocity difference suggesting the larger velocity dispersions in these multiply systems. The percent of 
the E-type galaxies is marked in Fig. 7 for each sample.
The higher fraction of the early type galaxies in CP1 and HDCG samples can be the evidences of the previous evolution/merging that occured in that systems.  As we did not apply the criteria of 
isolation for the selection of the compact pairs, we can assume that our CP1 systems are situated in the 
centres of the more populated groups and clusters. 

On the other hand, the closest double systems show less E-type galaxies that can tell us about the later evolution stage of 
these groups: convergence or outgoing merging (see for example Karachentsev 1987, Coziol et al. 2004, Melnyk et al. 2006). Patton et al. (2011) also found that the 
fraction of extremely blue galaxies is higher in close galaxy pairs than in the distant pairs. The 2MIG isolated galaxies, i.e. those galaxies that are located in the underdense 
regions without any significant enviromental influence, has the lowest percent of E-type galaxies. According to Croton and Farrar (2008), 
a population of passive elliptical galaxies in underdense regions could be naturally explained by an environment independent star formation shut-down mechanism.

\begin{figure*}
\tabcolsep 0 pt
\begin{tabular}{ccc}
\epsfig{file=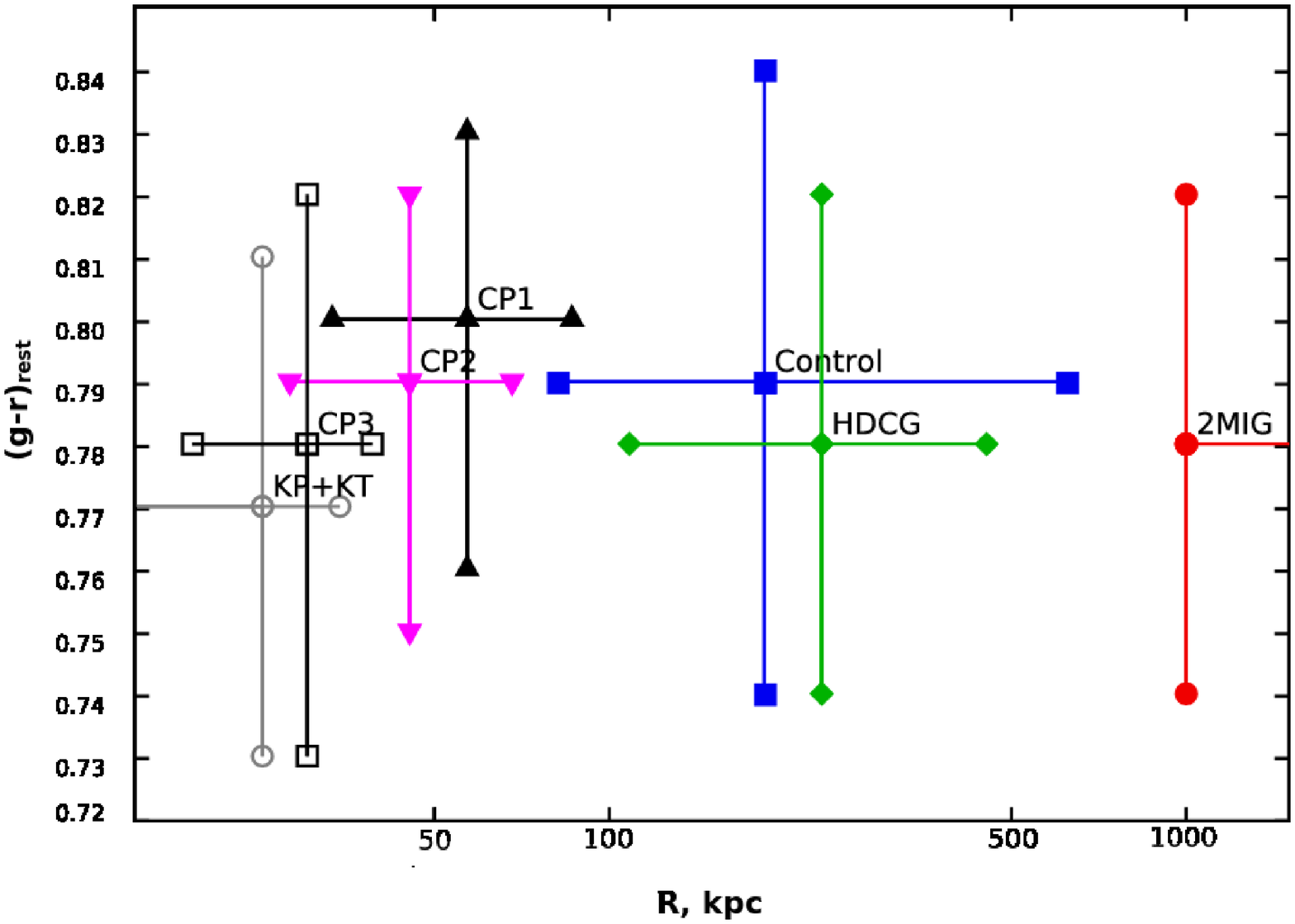,width=6.2cm,trim=0 0 40 0,clip} &
\epsfig{file=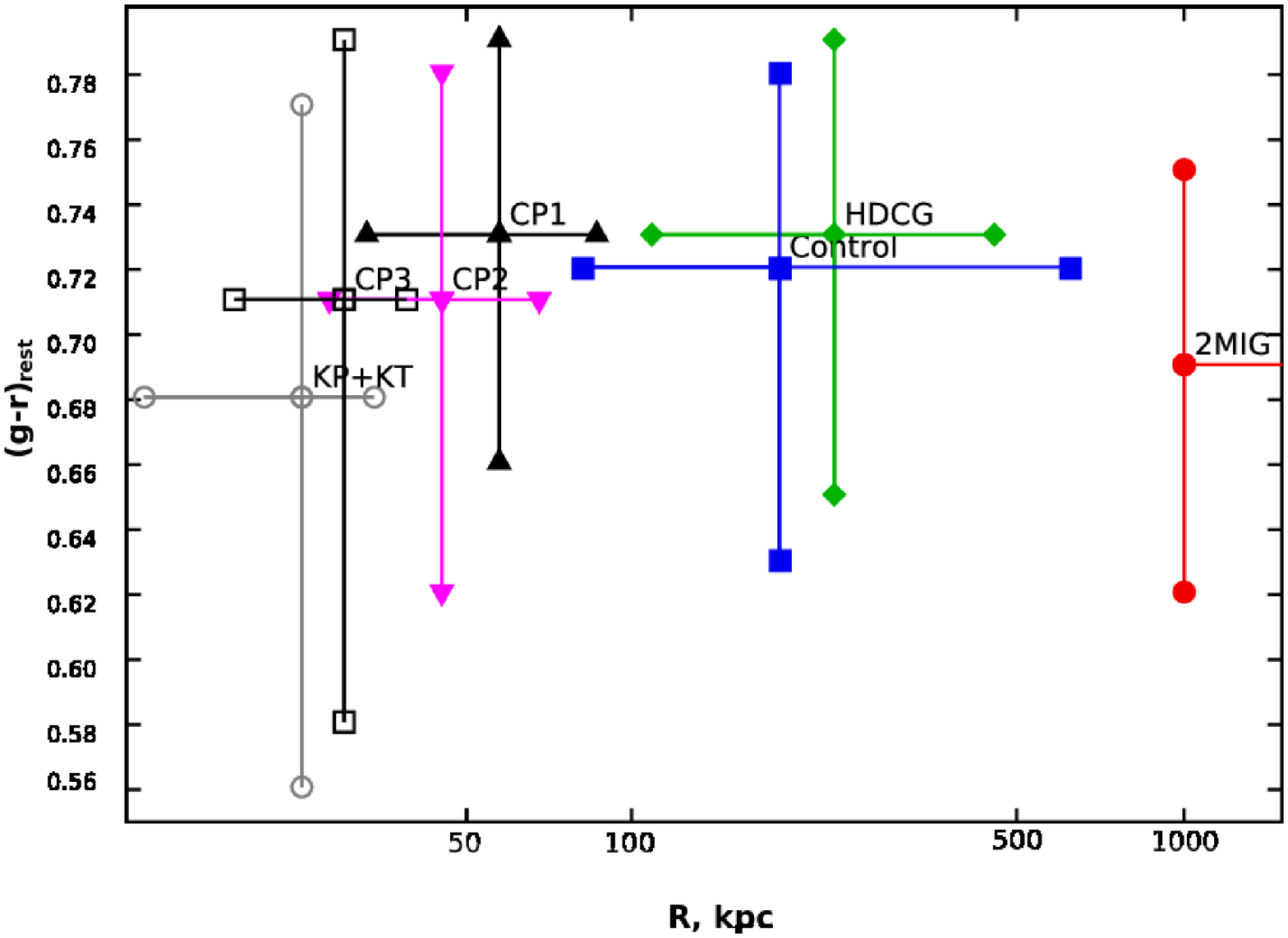,width=6.2cm,trim=0 0 40 0,clip} &
\epsfig{file=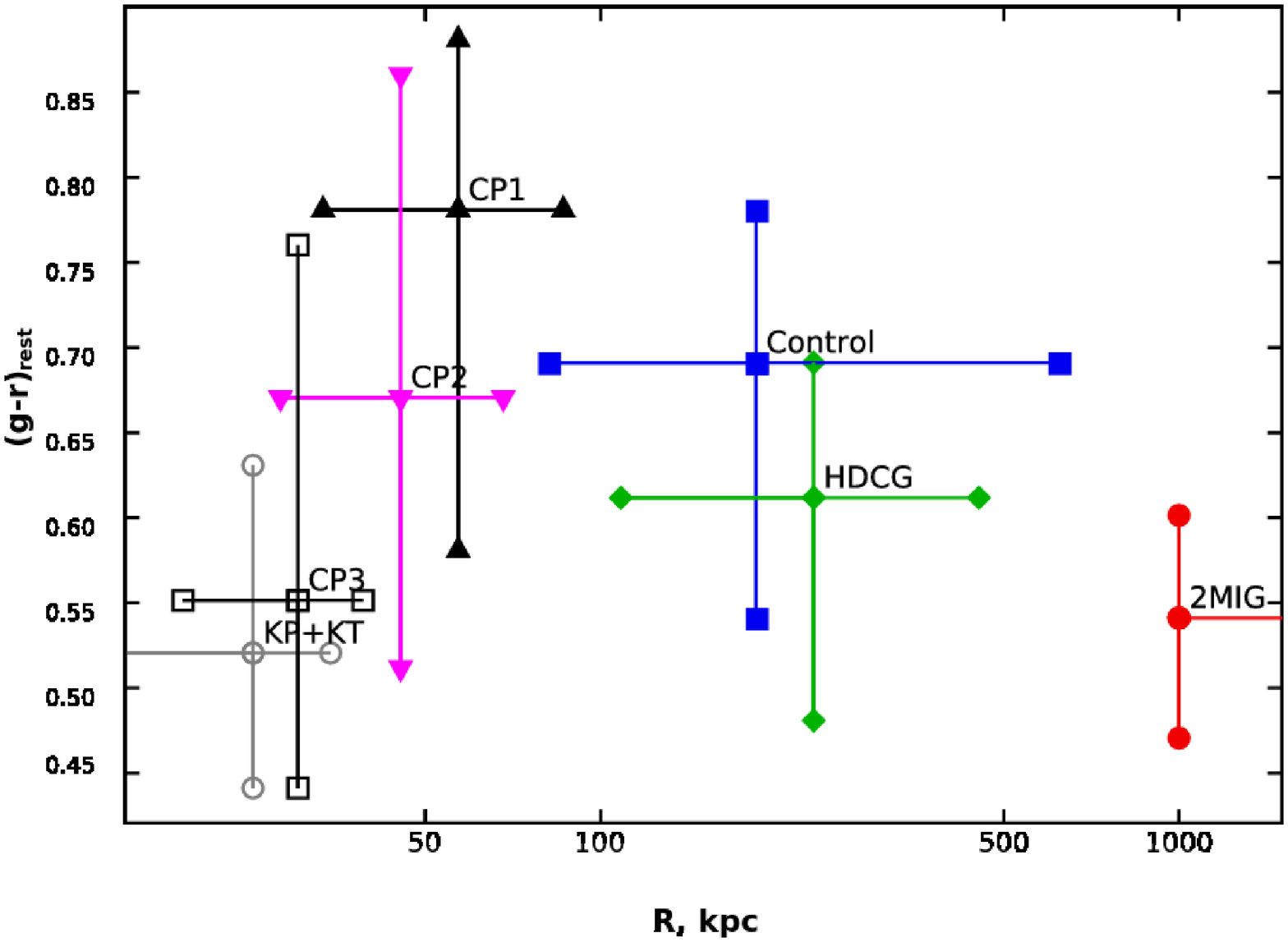,width=6.2cm,trim=0 0 40 0,clip} \\
E & M & L \\
\end{tabular}
\caption{ Dependence of $(g-r)_{rest}$ colour indices on the projected linear distance $R$ between galaxies: medians, 1 and 3 quatiles are shown for each sample.}
\label{8}
\end{figure*}

Fig. 8 presents the dependence of $(g-r)_{rest}$ colour indices on the projected linear distance $R$ between galaxies. The only median and quartile values for each 
sample are shown. For all three E, M and L types we observe the same tendency: with decreasing the distance between galaxies the median value of the colour indices 
increasing forming a peak on CP1 sample and then, again, the median colour index is decreasing with minimum on the KP+KT sample. Meanwhile we know from Table 4 that 
the difference between $(g-r)_{rest}$ colours of E-type galaxies is significant only for the 2MIG and CP1 samples (the latter sample is redder). 
Here we are in general agreement with Patton et al. (2011) who did not find any difference for red galaxies at any separations. The similar conclusion was made also by
Fernandez Lorenzo et al. (2012).

For the M and L-type galaxies, the significant difference is observed for HDCG, CP1 and CP2 samples, which galaxies are significantly redder than 2MIGs. 
So we see that samples 
of multiply systems with large projected distance (velocity dispersions) contain galaxies with heavy bulges which contain dust and old stellar populations. Here we are in
the agreement with our finding from section 3 (M and L-type Control galaxies are also redder than 2MIGs) and other studies: Rojas et al. (2004), Hoyle et al. (2005), 
(2012), Sorrentino et al. (2006). Patton et al. (2011) also found that samples of pair galaxies had redder galaxies than their control sample. 
 Note, that ($J-H$)$_{rest}$ colours of galaxies in pair and multiply systems are also redder than 2MIG ones.

The closest pairs KP+KT and CP3 have similar $(g-r)_{rest}$ colour to 2MIGs. However M-types KP+KT galaxies are, on the contrary, significantly bluer than 2MIG galaxies. 
Here we are in the agreement with Fernandez Lorenzo et al. (2012),  who found that $(g-r)_{rest}$ colours of spiral isolated galaxies are slightly redder than those of 
close pairs 
because of the outgoing star formation in latter systems. Trinh et al. (2013) recently found that the galaxies in pairs have a significant red excess for ''red'' 
($g-r\geq $0.
68) galaxies and a weak blue excess for the ``blue'' ($g-r<$0.68) galaxies in comparison with isolated objects. In summary, we also found that galaxies in very 
close pairs 
tend to be bluer than the isolated galaxies but here we probably limited by $Ks$ selection which does not allow us to select bluer systems.

\section{Conclusions}

We have considered the near-infrared (2MASS) and optical (SDSS) colours of isolated galaxies from the 2MASS XSC selected sample of isolated
galaxies (2MIG) in comparison to corresponding colours for 
objects selected from other denser environments by different selection criteria. In general, our results show that the colours of galaxies are strongly depend on 
the external factors. Our main findings are as follows:

(1)  Early (E) type galaxies show similar $(J-H)_{rest}$ and $(g-r)_{rest}$ colours in all types of environment. Exception is for the massive early type galaxies located 
in pairs with high velocity difference ($\sim$180 km/s), which are significantly redder and brighter than isolated galaxies. We assume that
these pairs are located in the centres of more populated groups and clusters.

(2)  M and L types of 2MIG galaxies which have faint neigbours are brighter and redder than those without faint neighbours. Moreover, randomly taken 
galaxies from the field and group/clusters environments with the same absolute magnitude $M_{Ks}$ as 2MIG galaxies of given morphological 
types, are also redder in their $(g-r)_{rest}$ colours than isolated galaxies.

(3) Reported in (2) tendency is much strongly appeared for the objects taken from much denser environment in comparison with isolated galaxies. In general, galaxies of 
morphological types T$>$1 from groups and triplets/pairs have redder near-infrared colours $(J-H)_{rest}$ than 2MIG 
isolated galaxies. The $(g-r)_{rest}$ colours of galaxies from groups and pairs with high velocity difference ($dV\sim$180 km/s; projected radius R$\sim$60 kpc) 
are also significantly 
redder than corresponding colours of 2MIG galaxies. On the contrary, the members of most compact pairs ($dV\sim$50 km/s, $R\sim$30 kpc) show the same colour 
and even tend to be bluer than 2MIG galaxies.

(4) The fraction of the early type galaxies in groups and pairs is 2 times higher than in isolated galaxies (48-62\% vs. 31\%).

\section*{Acknowledgments}

We are grateful to Igor Karachentsev for fruitful discussions and suggestions. We are thankful to Ross Parkin for his useful remarks that improved the paper. 
We would like also to thank the anonymous reviewer for his/her the careful reading of the paper and helpful comments. 
This research has made use of the NASA/IPAC Extragalactic Database (NED) which is operated by the Jet Propulsion Laboratory, 
California Institute of Technology, under contract with the National Aeronautics and Space Administration. We acknowledge the usage 
of the HyperLeda database (http://leda.univ-lyon1.fr). In our work we have also used SDSS-III data, funding for which has been provided by 
the Alfred P. Sloan Foundation, the Participating Institutions, the National Science Foundation, and the U.S. Department of Energy Office of Science. 
The SDSS-III web site is http://www.sdss3.org/. This work was supported by Ukrainian SFFR grant F53.2/15, grants of the Ministry of Education and Science of 
the Russian Federation N8523 and RFFR13-02-90407.

\label{lastpage}


\begin{thebibliography}{}

\bibitem {} Aihara H. et al. 2011, ApJS, 193, id. 29
\bibitem {} Allam S. S., Tucker D. L., Lee B. C., Smith J. A. 2005, AJ, 129, 2062
\bibitem {} Anderson M. E.,  Bregman J. N., Dai X. 2013, ApJ, 762, id. 106
\bibitem {} Bamford et al. 2009, MNRAS, 393, 1324 
\bibitem {} Blanton M. R., Eisenstein D., Hogg D. W., Schlegel D. J., Brinkmann J. 2005, ApJ, 629, 143
\bibitem {}  Blanton M. R., Moustakas J. 2009, ARA\&A, 47, 159
\bibitem {} Coziol R., Torres-Papaqui J. P, Plauchu-Frayn I., Islas-Islas J. M., Ortega-Minakata R. A.,  Neri-Larios D. M.,  Andernach H. 2011, Revista Mexicana de Astronomia y Astrofisica, 47, 361
\bibitem {} Coziol R., Brinks E., Bravo-Alfaro H. 2004, AJ, 128, 68
\bibitem {} Crook A. C., Huchra J. P., Martimbeau N., Masters K. L., Jarrett T., Macri L. M., 2007, ApJ, 655, 790
\bibitem {} Croton D. J., Farrar G.R., 2008, MNRAS, 386, 2285
\bibitem {} de Vaucouleurs G. 1971 . PASP, 1971, 83, 113
\bibitem {} Deng X. F., Yang B., Luo C. H., Qian X. X., Ding Y. P. 2012, Revista Mexicana de Astronomia y Astrofisica 48, 293
\bibitem {}  Durbala A.,  Sulentic J. W.,  Buta R., Verdes-Montenegro L. 2008, MNRAS 390, 881
\bibitem {} Elyiv A., Melnyk O., Vavilova I. 2009, MNRAS, 394, 1409
\bibitem [Elyiv et al.(2013)]{2013AstBu..68....1E} Elyiv, A.~A., Karachentsev, I.~D., Karachentseva, V.~E., Melnyk, O.~V., \& Makarov, D.~I.\ 2013, Astrophysical Bulletin, 68, 1
\bibitem {} Fernandez Lorenzo M. , Sulentic J. , Verdes-Montenegro L., Ruiz J. E., Sabater J., Sanchez S. 2012,  A\&A ,540, id.A47 
\bibitem {} Hernandez-Toledo H. M., J. A. Vazquez-Mata, L. A. Martinez-Vazquez, Choi Y.-Y., Park C. 2010, AJ, 139, 2525
\bibitem {} Hoyle F., Randall R. R., Michael S. V., Brinkmann J. 2005, ApJ, 620, 618
\bibitem {} Hoyle F., Vogeley M.S., Pan D., 2012, MNRAS, 426, 3041
\bibitem {} Jarrett T. N., Chester T., Cutri R., Schneider S., Skrutskie M., Huchra J. P. 2000, AJ, 119, 2498
\bibitem {} Karachentsev I. D.  Double galaxies. Moscow, Izdatel'stvo Nauka, 1987, 280 p.
\bibitem {} Karachentsev I. D., Makarov D. I., Karachentseva V. E., Melnyk O. V. 2011, Astrophys. Bull., 66, 1
\bibitem {} Karachentsev I. D., Karachentseva V. E., Melnyk O. V., Elyiv A. A., Makarov D. I 2012, Astrophys. Bull., 67, 353
\bibitem {} Karachentseva V. E. 1973, Soobshch. Spets. Astrofiz. Obs., 8, 3
\bibitem {} Karachentseva V.E., 1980, Soviet Astronomy, 24, 665
\bibitem {} Karachentseva V. E., Karachentsev I. D.,  Shcherbanovsky A. L. 1979, Astrof. Issledovaniia, 11, 3
\bibitem {} Karachentseva V.E., Karachentsev I.D. 2000, Astronomy Reports, 44, 501
\bibitem {} Karachentseva V. E., Mitronova S. N., Melnyk O. V., Karachentsev I. D. 2010, Astrophys. Bull., 65, 1
\bibitem {} Karachentseva V. E., Karachentsev I. D., Sharina M. E. 2010a, Astrophysics, 53, 462
\bibitem {} Karachentseva V. E., Karachentsev I. D., Melnyk O. V. 2011, Astrophys. Bull., 66, 389
\bibitem {} Kreckel K., Joung M. R., Cen R. 2011, ApJ, 735, 132
\bibitem {} Kreckel K., E. Platen E., Aragon-Calvo M. A., van Gorkom J. H., van de Weygaert R., van der Hulst J. M., Beygu B. 2012, AJ, id. 16
\bibitem {} Kudrya Yu. N., Karachentseva V. E., Karachentsev I. D., 2011, Astrophysics, 54, 445
\bibitem {} Kudrya Yu. N.,  Karachentseva V. E. 2012, Astrophysics, 55, 435
\bibitem {} Leon S. et al. 2008, A\&A, 485, 475 
\bibitem {} Lisenfeld U. et al. 2007, A\&A 462, 507 
\bibitem {} Lisenfeld U. et al. 2011, A\&A 534, id.A102
\bibitem {} Makarov D., Karachentsev I. 2011, MNRAS, 412, 2498
\bibitem {} Masters K. L., Giovanelli R., Haynes M. P. 2003, AJ, 127, 1257
\bibitem {} Masters K. L. et al. 2010, 404, 792
\bibitem {} Melnyk O. V. 2006, Astronomy Letters, 32, 302 
\bibitem {} Patiri S. G., Prada F., Holtzman J., Klypin A., Betancort-Rijo J.  2006, MNRAS, 372, 1710 
\bibitem {} Patton D.R., Ellison S.L., Simard L., McConnachie A. W., Mendel J. T. 2011,  MNRAS, 412, 591
\bibitem {} Paturel G., Petit C., Prugniel P., Theureau G., Rousseau J., Brouty M., Dubois P., Cambr{\'e}sy L., 2003, A\&A, 412, 45
\bibitem {} Poggianti B. 1997, A\&AS, 122, 399
\bibitem {} Rojas R. R., Vogeley M. S., Hoyle F., Brinkmann J. 2004, AJ, 617, 50
\bibitem {} Sabater J., Leon S., Verdes-Montenegro L., Lisenfeld U., Sulentic J., Verley S., 2008, A\&A, 486, 73 
\bibitem {} Sabater J., Verdes-Montenegro L., Leon S., Best P., Sulentic J., 2012, A\&A, 545, id.A15
\bibitem {} Sorrentino G., Antonuccio-Delogu V., Rifatto A. 2006, A\&A 460, 673
\bibitem {} Strateva I. et al. 2001, AJ, 122, 1861
\bibitem {}  Trinh C.T., Barton E.J., Bullock, James S., Zentner A. R., Wechsler R. H. 2013,  MNRAS 1301.5870
\bibitem {} Turner E.L., Gott J.R. 1977, ApJ, 197, L89
\bibitem {} Varela J., Moles M., Marquez I., Galletta G., Masegosa J., Bettoni D. 2004, A\&A, 420, 873
\bibitem {} van der Wel A. 2008, ApJ, 675, L13
\bibitem {} Vavilova I. B., Melnyk O. V., Elyiv A. A. 2009, AN, 330, 1004
\bibitem {} Verley S. et al. 2007, A\&A, 472, 121
\bibitem {} von Benda-Beckmann A. M. \& Muller V. 2008, MNRAS, 384, 1189 
\bibitem {} Zwicky F., Herzog E., Wild P., Karpowicz M., Kowal C. CGCG, 1961, , Pasadena: California Institute of Technology


\end{thebibliography}
\end{document}